\documentclass[reprint,aps,prl,amsmath,amssymb,superscriptaddress]{revtex4-2}
\usepackage{graphicx}
\usepackage{xcolor}
\usepackage[colorlinks,linkcolor=blue!50!black,citecolor=blue!50!black,urlcolor=blue!50!black]{hyperref}

\begin{document}

\title{Stationary superfluid wakes above the critical velocity}

\author{Niccol\`o Geracitano}
\affiliation{Gran Sasso Science Institute, Viale Francesco Crispi 7, 67100, L'Aquila, Italy}
\author{Francesco Viola}
\affiliation{Gran Sasso Science Institute, Viale Francesco Crispi 7, 67100, L'Aquila, Italy}
\affiliation{INFN-Laboratori Nazionali del Gran Sasso, Assergi, Italy}
\author{Vincenzo Citro}
\affiliation{DIIN, Universit\`a di Salerno, Via Giovanni Paolo II 132, 84084, Fisciano, Italy}
\affiliation{CNR-Nanotec, Unit\`a di Roma, Piazzale Aldo Moro 5, 00185 Rome, Italy}

\date{\today}

\begin{abstract}
The classical picture of superfluid flow past an obstacle associates
the onset of sustained vortex shedding with the loss of the stationary
vortex-free state. Here, we show that stationary wakes persist beyond
the shedding threshold up to a higher critical velocity. Through numerical
simulations combined with numerical continuation of the two-dimensional
Gross-Pitaevskii equation, we identify previously unexplored branches of stationary
solutions at subsonic supercritical velocities. These branches include
wakes with a vortex-antivortex pair at rest behind the obstacle, forming
a quantum counterpart of the viscous recirculation bubble.
An
analytical theory that accurately predicts the pair's equilibrium position and antisymmetric instability is further presented, connecting the higher critical velocity
to the physical properties of the solutions. Building on the discovery of these new states, we develop a feedback protocol designed for experimental implementation and demonstrate in time-dependent simulations that it suppresses vortex shedding above the vortex-nucleation threshold.
\end{abstract}

\maketitle

A uniform superfluid flow is dissipationless below the Landau critical
velocity, above which elementary excitations can be created~\cite{Landau41}.
For a uniform weakly interacting Bose gas, this
velocity is the speed of sound $c$~\cite{Bogoliubov47}, whereas an obstacle immersed in the fluid accelerates the flow at its sides
and lowers this threshold. Using direct numerical simulations of the
Gross-Pitaevskii (GP) equation, Frisch, Pomeau, and Rica~\cite{FPR92}
showed that a stationary vortex-free solution  exists only up to a critical
velocity $v_{c1}<c$, at which the flow at the sides of the obstacle
becomes locally sonic. Below this threshold, the obstacle feels no
drag. Subsequent studies showed that the vortex-free
flow merges in a saddle-node bifurcation with an unstable stationary
flow containing a vortex pair beside the
obstacle at $v_{c1}$~\cite{HB00,Rica01,PNB05}, while above this value vortex pairs
are shed periodically and the fluid starts exerting a drag on the
obstacle~\cite{FPR92,WMA99}. The onset of shedding was recently shown
to occur through a global saddle-node on an invariant circle (SNIC)
bifurcation~\cite{GVLC26}, characterized by an emission period that
diverges as $(v_\infty-v_{c1})^{-1/2}$, in contrast with viscous cylinder
wakes, in which shedding starts through a local Hopf
bifurcation~\cite{Williamson95,Fabre19}. In the latter, the
stationary wake persists above the instability threshold and provides
the base flow for wake stability
analysis~\cite{Fornberg80,Jackson87,Fornberg85,Barkley06,GL07}.
For superfluid wakes, instead, no stationary two-dimensional solution
has so far been identified at subsonic inlet velocities above
$v_{c1}$ and it remains unclear whether the shedding threshold
also marks the existence limit of stationary flows. Identifying such
solutions would be crucial to provide reference states for studying
supercritical wake instabilities and targets for feedback control
of vortex shedding.

This Letter shows that such supercritical stationary wakes exist. We identify
a branch of stationary GP solutions that extends to a second critical
velocity $v_{c2}>v_{c1}$ and contains a vortex-antivortex pair in
hydrodynamic equilibrium behind the obstacle. The reversed flow between
the obstacle and the pair forms a quantum analog of the classical
recirculation bubble, and building on this correspondence we show how classical
point-vortex theory can predict the pair's
equilibrium position and the growth rate of its antisymmetric
instability as functions of the inlet velocity. We further present a
feedback protocol on the obstacle position that successfully sustains these otherwise unstable
wakes, providing the first
example of non-shedding superfluid flows in supercritical
regimes.

\begin{figure}[t]
\centering
\includegraphics[width=\columnwidth]{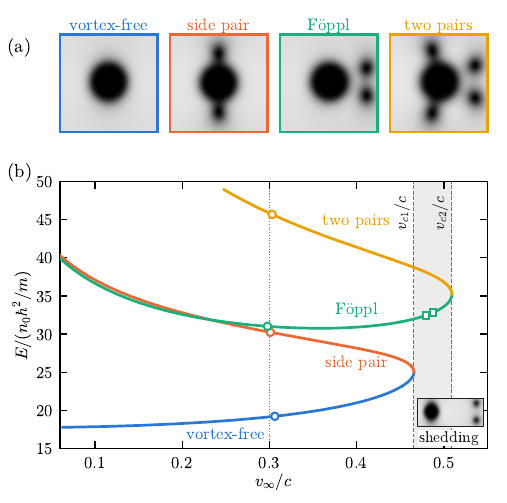}
\caption{Branches of stationary solutions. (a) Density $|\psi|^2/n_0$ of one solution of each branch at
$v_\infty/c\approx0.30$, in a window of side $20\,\xi$ centered on the obstacle. The flow comes from the left.
Black: low density regions (obstacle and vortex cores). (b) Energy of the flow. Circles mark the solutions shown
in (a). The vortex-free and side-pair branches meet at $v_{c1}$, the F\"oppl and two-pair branches at $v_{c2}$.
In the shaded band only the F\"oppl and two-pair branches exist. Inset: density of a
time-dependent solution at $v_\infty/c=0.470$, where the F\"oppl branch coexists with periodic shedding,
showing the obstacle and a vortex pair shed downstream.}
\label{fig:branches}
\end{figure}

We consider a two-dimensional quantum fluid of density $n_0$ that flows with velocity $v_\infty$ along $x$ in the frame of
the obstacle~\cite{GVLC26,GVLC26b}. A stationary state satisfies
\begin{equation}
-\frac{\hbar^2}{2m}\nabla^2\psi+V\psi+g\left(|\psi|^2-n_0\right)\psi-\frac{mv_\infty^2}{2}\,\psi=0,
\label{eq:gp}
\end{equation}
with $\psi\to\sqrt{n_0}\,e^{imv_\infty x/\hbar}$ far from the obstacle. The superfluid velocity is
$\mathbf v=(\hbar/m)\nabla\arg\psi$, the speed of sound is $c=(gn_0/m)^{1/2}$, and the healing length is
$\xi=\hbar/(2mgn_0)^{1/2}$. The obstacle is the Gaussian potential $V=V_0e^{-r^2/d^2}$ with $V_0=100\,gn_0$ and
$d=1.62\,\xi$, acting as an effectively impenetrable barrier of radius
$a\simeq4.29\,\xi$~\cite{SM}.
We discretize Eq.~\eqref{eq:gp} with second-order finite differences on a uniform grid of spacing $h=\xi/4$ in the
square $[-L,L]^2$, with $L$ between $256\,\xi$ and $512\,\xi$.
Branches of solutions are followed in $v_\infty$ by pseudo-arclength continuation, following the approach 
described in Ref.~\cite{GVLC26b}. This method allows the identification of the four 
branches shown in Fig.~\ref{fig:branches}, which we name after the features of the 
vortices they contain. The vortex-free branch is the classical solution reported in Refs.~\cite{FPR92,HB00,PNB05}.
The side-pair branch has a vortex pair on the line $x=0$, with the vortex of negative circulation in the upper
half-plane. This is the nucleation solution already described in Refs.~\cite{HB00,PNB05}, and it meets the vortex-free
branch in a fold at $v_{c1}=0.466\,c$. The novel branch, indicated by the name F\"oppl for reasons that will be made clear later, has a vortex-antivortex pair behind the
obstacle, at $(x_0,\pm y_0)$ with $x_0>0$, and it is connected with a fold at $v_{c2}=0.509\,c$ to a
two-pairs branch. Figure~\ref{fig:branches}(b) shows the energy of the solutions
$E=\int\bigl[(\hbar^2/2m)|\nabla\phi|^2+V|\phi|^2+(g/2)(|\phi|^2-n_0)^2\bigr]\,d^2x$ with 
$\phi=\psi\,e^{-imv_\infty x/\hbar}$.
In the range in which the four solutions coexist, the vortex-free branch has the lowest energy and the two-pair branch
the highest.

Viscous wake dynamics provides a useful guide to characterize 
these solutions.
A simple theoretical model of the recirculation bubble was developed by F\"oppl in 1913, showing that two point vortices of opposite circulation
can stand behind a solid cylinder of radius $a$ placed in a uniform incompressible potential inviscid flow~\cite{Foppl13}.
In this setting, the position of the pair can be computed exactly.
In complex notation, $z=x+iy$, the circle theorem~\cite{MT68} gives the complex potential of a uniform stream past the
cylinder with a vortex of circulation $-\Gamma$ at $z_1=x_0+iy_0$ and a vortex of circulation $+\Gamma$ at $\bar z_1$,
\begin{equation}
w(z)=v_\infty\Bigl(z+\frac{a^2}{z}\Bigr)+\frac{i\Gamma}{2\pi}\ln\frac{z-z_1}{z-\bar z_1}
-\frac{i\Gamma}{2\pi}\ln\frac{z-a^2/\bar z_1}{z-a^2/z_1},
\label{eq:w}
\end{equation}
where the last term describes the contribution of the images of the two vortices located inside the cylinder. Each vortex thus moves with the velocity induced
at its position by the rest of the flow. Hydrodynamic equilibrium requires this velocity to vanish at $z_1$, resulting in two conditions,
which are linear and homogeneous in $v_\infty$ and $\Gamma$. For $x_0\neq0$ they reduce to~\cite{Foppl13,SM}
\begin{equation}
r_0-\frac{a^2}{r_0}=2y_0,\qquad \Gamma=4\pi v_\infty y_0\Bigl(1-\frac{a^4}{r_0^4}\Bigr),
\label{eq:foppl}
\end{equation}
with $r_0=|z_1|$. The first relation defines a curve that starts at the rear stagnation point of the cylinder and
approaches the $30^\circ$ angle from the downstream axis at large distances. In a classical fluid $\Gamma/v_\infty a$
can take any value, and each selects one point of the curve according to the second relation. 
In quantum fluids, instead, a singly quantized vortex has circulation $\Gamma=2\pi\hbar/m$, yielding
\begin{equation}
y_0\Bigl(1-\frac{a^4}{r_0^4}\Bigr)=\frac{\hbar}{2mv_\infty}.
\label{eq:quantized}
\end{equation}
The flow speed alone thus fixes the position of the pair. For $x_0=0$, the equilibrium condition on the transverse velocity is satisfied identically
while the streamwise one yields a second family of equilibria, with the vortices
 beside the cylinder. This family, identified but dismissed by F\"oppl as 
 experimentally unobserved in the classical case~\cite{Foppl13,SM}, corresponds
 to the quantum side-pair branch.

\begin{figure}[t]
\centering
\includegraphics[width=\columnwidth]{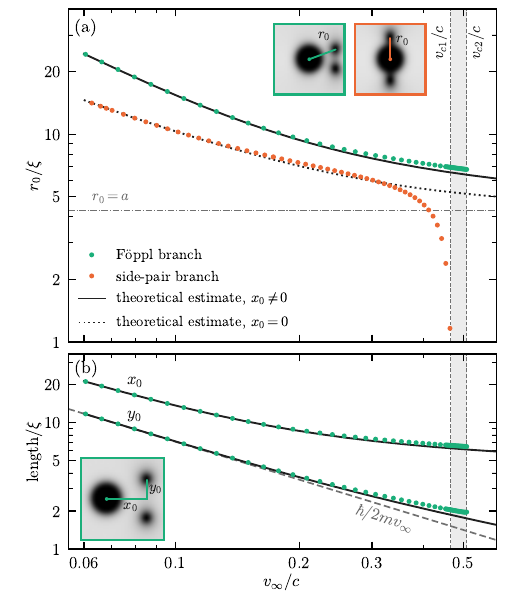}
\caption{Theoretical prediction \eqref{eq:foppl} against numerical results. (a) Distance $r_0$ of the upper vortex from the center of the obstacle on the F\"oppl branch (green) and on
the side-pair branch (orange). Solid line: Eq.~\eqref{eq:foppl} for $a=4.29\,\xi$ and
$\Gamma=2\pi\hbar/m$. Dotted line: the equilibrium of the same pair on the line $x_0=0$. Dash-dotted line: $r_0=a$.
(b) Coordinates $x_0$ and $y_0$ of the upper vortex of the F\"oppl
branch (circles), compared with Eq.~\eqref{eq:foppl} (solid lines) and $\hbar/2mv_{\infty}$ (dashed line).}
\label{fig:positions}
\end{figure}

Figure~\ref{fig:positions}(a) compares the distance $r_0$ of the upper vortex from the center of the obstacle with the theoretical prediction \eqref{eq:foppl} for the different
equilibria. In the $x_0\ne 0$ case, $r_0$
agrees with Eq.~\eqref{eq:foppl} within 1\% for $0.06\le v_\infty/c\le0.16$, where the pair is between $2.5\,a$ and $5.6\,a$ from the center.
For $x_0=0$, the side-pair branch agrees with the corresponding theoretical prediction within 1\% up to
$v_\infty/c=0.11$ and within 5\% up to $v_\infty/c\approx0.35$. Figure~\ref{fig:positions}(b) shows the accuracy of both formulas in Eq.~\eqref{eq:foppl} in predicting the actual positions of the vortex cores.
\textcolor{black}{For high values of $r_0$,} Eq.~\eqref{eq:quantized} gives
$y_0\simeq\hbar/(2mv_\infty)$, so the pair's upstream motion due
to mutual induction balances the inlet flow, while the obstacle
fixes the angular positions of the vortices through the first
relation in Eq.~\eqref{eq:foppl}. 
The accuracy of the theoretical model for the $x_0=0$ branch is limited by
the absorption of the dipole inside the depleted region of the obstacle for inlet velocities near $v_{c1}$, while 
near $v_{c2}$
the vortex cores of the quantum F\"oppl pair nearly touch ($2y_0\approx4\,\xi$) and compressibility and obstacle deformation effects are non-negligible. The validity of F\"oppl's theory for a quantum fluid, as opposed to its only
qualitative value in the viscous case, is a consequence of the fact
that, as already noted by Onsager~\cite{Onsager49,ES06}, the original assumptions of the point-vortex model
hold more closely in a superfluid than in a classical one: the vorticity is concentrated on
cores of size $\xi$, there is no viscosity, and the flow is potential \textcolor{black}{outside} vortex cores.

\begin{figure*}[t]
\centering
\includegraphics[width=\textwidth]{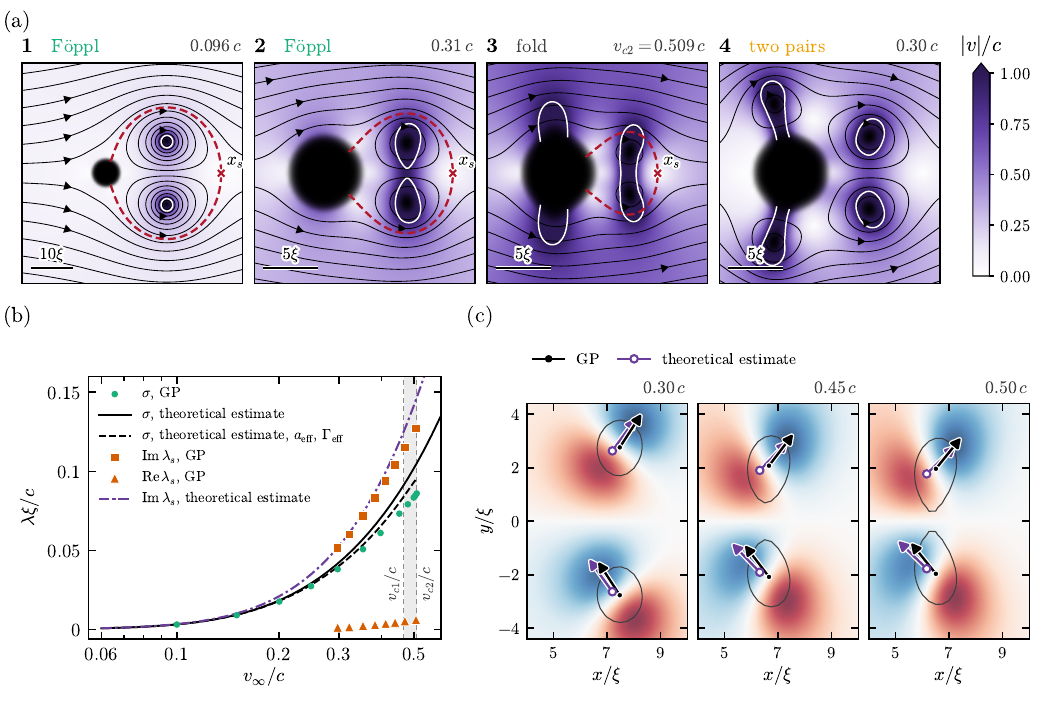}
\caption{(a) Representative solutions. Color: speed $|\mathbf v|/c$, darkened at obstacle and vortex cores, where the density vanishes.
White lines represent isocontours $|\mathbf v|=c$, while black lines are streamlines of the mass current.
We show the F\"oppl and two-pair branches,
which meet at $v_{c2}$ (panel 3). Red dashed lines in panels 1-3 represent the
streamline that bounds the recirculating region, which meets the axis at the stagnation point $x_s$ (red cross).
(b) Growth rate $\sigma$ of the unstable antisymmetric mode and symmetric
eigenvalue $\lambda_s$ of the quantum F\"oppl branch, compared with F\"oppl's
point-vortex pair for $a=4.29\,\xi$ and $\Gamma=2\pi\hbar/m$, and for
$a_{\text{eff}}$ and $\Gamma_{\text{eff}}$. (c) The antisymmetric unstable mode at $v_\infty/c=0.30$, 0.45, and
0.50. Color: density perturbation $\delta n=2\,\mathrm{Re}(\psi_0^*\delta\psi)$ (red positive, blue negative), in
arbitrary units. Thin lines represent the isocontours $|\psi_0|^2=n_0/4$ near the vortex cores. Arrows: direction of the displacement of
the cores in the Bogoliubov (black, from the GP cores, dots) and F\"oppl's unstable mode for $a=4.29\,\xi$ and
$\Gamma=2\pi\hbar/m$ (purple, from the point-vortex positions, open circles).}
\label{fig:flow}
\end{figure*}

Figure~\ref{fig:flow}(a) shows a sequence of representative states. On the F\"oppl branch the flow on the symmetry
axis behind the obstacle runs towards the obstacle, up to a stagnation point at $x=x_s$, beyond which it runs
downstream again. For $v_\infty=v_{c2}$ and $v_\infty/c=0.31$ and 0.096 we find $x_s=8.99\,\xi$, $11.5\,\xi$, and $27.0\,\xi$ from flow analysis, in good
 agreement with the values computed using Eq.~\eqref{eq:w}. Closed streamlines around each vortex define the two
counter-rotating cells of a separated wake.
The quantum bubble differs from the classical one in two respects. Its vorticity is concentrated in two quantized
cores, so the dimensionless strength of the eddies, $\Gamma/v_\infty a$, decreases as the flow speed increases. The
bubble therefore grows as the flow slows down, whereas the classical bubble grows with the Reynolds number. The quantum
bubble also exerts no drag, as
\textcolor{black}{the force on the obstacle, $F_x=\int|\psi|^2\partial_xV\,d^2x$, is
$|F_x|\approx4\times10^{-7}\,gn_0^2\xi$ at $v_\infty/c=0.48$ and $4\times10^{-6}\,gn_0^2\xi$ at $v_\infty/c=0.096$.}

The offset between the two folds can be explained by the structural properties of the solutions. Neglecting
the quantum pressure and the obstacle potential, the stationary flow obeys Bernoulli's law,
$g(n-n_0)+mv^2/2=mv_\infty^2/2$. The current $j=nv$ through a streamtube has a
maximum,
\begin{equation}
j^*=\frac{n_0}{c^2}\left(\frac{2c^2+v_\infty^2}{3}\right)^{3/2},
\label{eq:jstar}
\end{equation}
reached where the flow is locally sonic.
When the largest current through the line $x=0$ above the obstacle, $j_\mathrm{pole}$, reaches $j^*$, the
flow can no longer pass the obstacle in a stationary state and the branch folds back~\cite{FPR92,Rica01,Hakim97}.
At equal inlet speeds, however, the pole current is higher on the vortex-free branch than on the F\"oppl one.
According to the point-vortex model, the vortex pair and its images  at $v_{c2}$ modify the flow in the vicinity of the obstacle,
 lowering the largest speed on the surface of the cylinder by approximately 7\% and moving the maximum point to
$97^\circ$ from the downstream axis, which explains part of the velocity mismatch between the two folds and the position of vortex appearance in the two-pair branch.
Furthermore, vortex tracking techniques reveal that this transition occurs in the same way as on the vortex-free branch, with the generation of ghost vortices inside the obstacle~\cite{SM}.
The fold is thus a local event at the
poles, consistently with Landau's theory.

In classical fluids, F\"oppl's model predicts neutral stability to
perturbations that preserve the symmetry $y\to-y$ and instability
to antisymmetric ones~\cite{TA97,VMS11,EFM05}. We have computed the
Bogoliubov spectrum of the quantum F\"oppl branch for
$0.1\le v_\infty/c\le v_{c2}/c$, following the approach developed
in Ref.~\cite{GVLC26b}. At every sampled velocity, we find a pair
of real eigenvalues $\pm\sigma$ [Fig.~\ref{fig:flow}(b)] associated with an antisymmetric
mode, shown in Fig.~\ref{fig:flow}(c).
The spectrum also reveals a weaker symmetric instability associated with a complex pair of eigenvalues $\lambda_s$, $\overline{\lambda_s}$~\cite{SM}. Its energy decomposition is
consistent with a negative-energy vortex oscillation coupled to
positive-energy sound, suggesting a connection with the neutral
symmetric oscillation of F\"oppl's model.
Figure~\ref{fig:flow}(b) compares the antisymmetric growth rate
$\sigma$ with the point-vortex prediction, \textcolor{black}{showing good agreement at low velocities.}
In a neighborhood of $v_{c2}$, $\sigma$ is lower than the corresponding point-vortex estimate.
 This difference can be reduced by \textcolor{black}{defining}
 $a_{\text{eff}}$ and $\Gamma_{\text{eff}}$ in Eq.~\eqref{eq:foppl} to
  match the GP vortex positions $(x_0,y_0)$, since at higher velocities 
  the quantum F\"oppl pair lies further from the obstacle than the 
  theoretical prediction. The \textcolor{black}{effective} radius changes in the interval $[4.27\, \xi, 4.44\, \xi]$ over the whole velocity range, while the effective circulation exceeds
$2\pi\hbar/m$ by $0.5\%$ for $v_\infty/c=0.10$, growing to $16\%$ at
$v_{c2}$.
With the effective radius and circulation, $\sigma$ lies 2\% to 10\% below the point-vortex estimate~\cite{SM}.
As shown in Fig.~\ref{fig:flow} (c), the corresponding
 unstable mode is associated with a displacement of both cores to the same side
  of the axis. The lower vortex moves up and upstream, while
   the upper one moves downstream, so that the pair is shifted and turned.

\begin{figure}[t]
\centering
\includegraphics[width=\columnwidth]{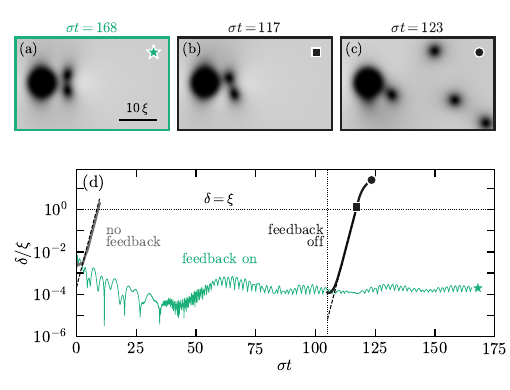}
\caption{Feedback control at $v_\infty=0.479\,c$.
(a)-(c) Density $|\psi|^2/n_0$ in a
$40\,\xi\times24\,\xi$ window.
(a) Feedback on, at the end of the simulation.
(b), (c) After feedback is switched off at $\sigma t=105$.
(d) Largest distance $\delta$ of the two vortex cores from their stationary positions relative to the obstacle,
with feedback on (green), without feedback after the same initial obstacle displacement (gray), and following
switch-off (black). Dashed lines indicate growth-rate $\sigma$. The symbols mark the times of (a)-(c).}
\label{fig:control}
\end{figure}

As is customary for classical systems, the analysis of unstable stationary 
solutions can be exploited to suppress shedding in supercritical regimes~\cite{GL07,SS90}.  In viscous wakes, F\"oppl's model
has guided the control of the steady
recirculation bubble by means of cylinder rotation,
although leaving out the
neutral symmetric eigenvalues~\cite{Protas04,Protas07}.
For the superfluid counterpart,
we integrate the time-dependent GP
equation starting from the F\"oppl state at two representative supercritical velocities
$v_\infty=0.479\,c$ and $0.488\,c$, and use the vortex-core positions to update a linear feedback
law, moving the obstacle in the cross-stream and in the 
streamwise direction to balance both the fast antisymmetric and
the slow symmetric eigenvalue. When the feedback is active, the cores remain within $7.8\times10^{-3}\,\xi$ of their stationary
positions throughout both runs, while the
obstacle displacement remains below $7.4\times10^{-3}\,\xi$
[Fig.~\ref{fig:control}]. After the initial transient, the
density profiles differ from the target F\"oppl density by less
than $7.5\times10^{-4}\,n_0$. Without  feedback, the same initial perturbation
produces a vortex displacement greater than one healing length for $t\gtrsim9/\sigma$, 
while with the control activated the state can be held without shedding.
Switching off feedback, releases the pair and the subsequent growth rate agrees with the one
computed by the Bogoliubov system to within 0.2\%. More details on the control design and the results are provided
in the End Matter.

In conclusion, we have shown that quantum flows past obstacles admit supercritical stationary solutions characterized by a vortex-antivortex pair
satisfying the equilibrium conditions derived by F\"oppl in 1913 for point vortices behind a solid cylinder. The quantization of
circulation fixes the position of the pair at each inlet speed, and the theoretical model, which has only a qualitative value for viscous
wakes, acquires predictive accuracy whenever quantum effects are negligible. The pair forms a drag-free recirculation bubble and, by slowing the flow
at the poles, delays the local sonic condition that ends the vortex-free flow at $v_{c1}$. The
speed at which a stationary branch ends thus depends on the structural properties of the solution, while the local criterion is the same. Like the theoretical
F\"oppl's solution, the quantum pair is unstable to antisymmetric displacements. Although unstable steady solutions can be observed experimentally only for short times without control, they contribute to the overall organization of
the dynamics, as together with their stable and unstable manifolds, they capture the coherent structures observed in experiments and simulations, while their bifurcations mark
qualitative changes of the flow~\cite{ATS27}. As an example, the fold at $v_{c1}$ marks the onset of vortex shedding, whose period diverges at threshold~\cite{GVLC26}.
Between $v_{c1}$ and $v_{c2}$, the F\"oppl and
two-pair states are the only stationary solutions we find. The Föppl flow therefore provides a natural reference state for describing the evolution of coherent structures in the supercritical quantum wake,
and its unstable mode is a candidate for the symmetry breaking observed
in the V-shaped wake~\cite{SSS10,SM}. Finally,  we have used the new branches to design a feedback control strategy acting on the obstacle position
that suppresses both the symmetric and antisymmetric instability,
 providing the first example of a subsonic non-shedding state above the critical velocity $v_{c1}$.

\begin{acknowledgments}
\end{acknowledgments}

\bibliography{bibliomain}

\onecolumngrid
\vspace{2\baselineskip}
\begin{center}\textbf{End Matter}\end{center}
\vspace{\baselineskip}
\twocolumngrid

\setcounter{equation}{0}
\renewcommand{\theequation}{A\arabic{equation}}
\renewcommand{\theHequation}{A\arabic{equation}}

We use the stationary supercritical ($v_\infty >v_{c1}$)
F\"oppl solutions at $v_\infty=0.479\,c$ and $0.488\,c$ as the
reference states for feedback, indicated by squares in Fig.~\ref{fig:branches}(b). We integrate the time-dependent GP
equation with the scheme of Refs.~\cite{GVLC26,GVLC26b}, using
$h=\xi/4$, $L=384\,\xi$, and a time step $\Delta t=0.12$ in units of $2mh^2/\hbar$.
Outgoing sound is absorbed in a layer of width
$50\,\xi$ at the inflow
and lateral boundaries, with a radiative condition at the outflow. Every 10 time steps, we locate the upper and lower vortex cores
$(x_\pm,y_\pm)$, and we define the antisymmetric (resp. symmetric) displacement relative to the equilibrium coordinates $(X+x_0,Y\pm y_0)$,
\begin{align}
\mathbf d_{\mathrm a}&=\Bigl(\tfrac{x_+-x_-}{2},\
                 \tfrac{y_++y_-}{2}-Y\Bigr),\nonumber\\
\mathbf d_{\mathrm s}&=\Bigl(\tfrac{x_++x_-}{2}-X-x_0,\
                 \tfrac{y_+-y_-}{2}-y_0\Bigr),
\label{eq:es}
\end{align}
where $(X,Y)$ is the obstacle position. The components of $\mathbf d_{\mathrm a}$ measure the tilting
and transverse displacement of the pair, while those of $\mathbf d_{\mathrm s}$
measure changes in its downstream position and separation, and
both vanish in the reference state. Obstacle motion is allowed in both the cross-stream and streamwise directions, 
with the former acting
on $\mathbf d_{\mathrm a}$ and the latter on $\mathbf d_{\mathrm s}$.
The feedback updates the obstacle velocity every $T_s=100\,\Delta t$,
keeping it constant between updates. The two responses decouple to linear order, allowing a separate
feedback law for each direction.

Unlike in the point-vortex
model~\cite{Protas04,Protas07}, the response of the GP flow to the
obstacle displacement is delayed,  as vortex cores first move
in the opposite direction and follow the obstacle only after about
$4\,\xi/c$. We therefore describe the response in each direction by a linear model with
memory, identified from the GP flow, and design the feedback on this model as a
linear-quadratic regulator (LQR)~\cite{AM90,HSSS12}. For the cross-stream direction,
we define $\mathbf d_k=\mathbf d_\mathrm{a}(kT_s)$ and $u_k$ the obstacle
velocity $\dot Y$ held during the $k$th interval, and we model the response of the flow as
\begin{equation}
\begin{aligned}
\mathbf d_{k+1}&=\sum_{i=0}^{2}A_i\,\mathbf d_{k-i}
               +\sum_{j=0}^{5}\mathbf b_j\,u_{k-j},\\
Y_{k+1}&=Y_k+T_s\,u_k .
\end{aligned}
\label{eq:arx}
\end{equation}
The six most recent velocities are kept to cover the response delay. 
For this model, the
feedback law
\textcolor{black}{
\begin{equation}
u_k=-\sum_{i=0}^{2}\mathbf K_i\cdot\mathbf d_{k-i}
    -\sum_{j=1}^{5}G_j\,u_{k-j}-G_Y\,Y_k ,
\label{eq:law}
\end{equation}
}
is the minimizer of the cost function
\begin{equation}
J=\sum_{k\ge0}\bigl(|\mathbf d_k|^2+q_Y\,Y_k^2+q_u\,u_k^2\bigr),
\label{eq:cost}
\end{equation}
with $q_Y=1$ and $q_u=78\,\xi^2/c^2$. The term depending on $Y$ prevents the
obstacle and the pair from drifting together, while the term in $u$ limits the
obstacle velocity.
The streamwise law follows in the same way
from $\mathbf d_\mathrm{s}$, $X$, and $\dot X$, with the same weights.

At $0.479\,c$, the cross-stream coefficients $A_i$, $\mathbf b_j$ are
fitted by least squares to the response to small obstacle displacements
and to a short controlled run in which a weak random binary
signal is added to the obstacle velocity, to separate the response of
the flow from the action of the controller. The streamwise coefficients
are fitted analogously using the response to streamwise displacements.
At $0.488\,c$ the calibration is repeated starting from the $0.479\,c$
controllers.
We test the design by comparing the evolution of the full GP equation with
and without these feedback laws. The maximum allowed obstacle velocity is
$0.057\,c$ and is never reached at both inlet velocities.

The controlled and uncontrolled
runs start from the same stationary F\"oppl solutions.
We perturb
the flow by displacing the obstacle to
$(X_0,Y_0)=(2.5,1.25)\times10^{-3}\,\xi$, leaving the wave
function initially unchanged. Without feedback, $|\mathbf d_\mathrm{a}|$
grows at a rate consistent with the Bogoliubov eigenvalue $\sigma$ and
reaches order $\xi$ after approximately $1.1\times 10^4$ time steps.
 At $0.479\,c$, after an initial transient,
feedback
maintains $|\mathbf d_\mathrm{a}|\le6.0\times10^{-4}\,\xi$ and
$|\mathbf d_\mathrm{s}|\le2.9\times10^{-4}\,\xi$ until the run ends.
The obstacle stays within
$5.8\times10^{-4}\,\xi$ of the origin, and
the near-wake density differs from the
stationary density by less than $6\times10^{-4}\,n_0$.
At $0.488\,c$, the state is held until the end of the simulation,
with
$|\mathbf d_\mathrm{a}|\le9.1\times10^{-4}\,\xi$, $|\mathbf d_\mathrm{s}|\le2.7\times10^{-4}\,\xi$, and near-wake density deviations below $7.5\times10^{-4}\,n_0$ after the
 transient. Obstacle motion in both directions is needed, as cross-stream feedback alone holds
$\mathbf d_\mathrm{a}$,
but the symmetric oscillation grows with a rate consistent with
$\mathrm{Re}\,\lambda_s$.
We have checked that the illustrated procedure survives initial obstacle displacements
up to $(X_0,Y_0)=(5,5)\times10^{-3}\,\xi$ combined with a random
perturbation of $\psi$ of maximum amplitude
$2\times10^{-3}\sqrt{n_0}$.
When both feedback laws are switched off,
 the pair escapes at the antisymmetric growth
rate $\sigma$ and reaches $|\mathbf d_\mathrm{a}|=\xi$
after approximately $1.4 \times 10^4$ time steps, as shown in
Fig.~\ref{fig:control} and in Supplemental Movie 1~\cite{SM}.

\end{document}


\title{Supplemental Material for ``Stationary superfluid wakes above the critical velocity''}

\author{Niccol\`o Geracitano}
\affiliation{Gran Sasso Science Institute, Viale Francesco Crispi 7, 67100, L'Aquila, Italy}
\author{Francesco Viola}
\affiliation{Gran Sasso Science Institute, Viale Francesco Crispi 7, 67100, L'Aquila, Italy}
\affiliation{INFN--Laboratori Nazionali del Gran Sasso, Assergi, Italy}
\author{Vincenzo Citro}
\affiliation{DIIN, Universit\`a di Salerno, Via Giovanni Paolo II 132, 84084, Fisciano, Italy}
\affiliation{CNR--Nanotec, Unit\`a di Roma, Piazzale Aldo Moro 5, 00185 Rome, Italy}

\maketitle

\section{Numerical setup}
\label{sec:methods}

The stationary Gross--Pitaevskii (GP) equation~(1) is discretized with second-order finite differences on a
uniform grid of spacing $h=\xi/4$ in $[-L,L]^2$. The standard domain has $L=384\,\xi$, with
$\psi=\sqrt{n_0}\,e^{ikx}$ on its boundary and $k=mv_\infty/\hbar$.
Stationary branches are followed by pseudo-arclength continuation, and their stability is obtained by linearizing
the same discrete equations. Details of the continuation, eigensolvers, and time-integration scheme are given in
Refs.~\cite{GVLC26,GVLC26b}.

The time-dependent equation is
\begin{equation}
i\hbar\,\partial_t\psi=-\frac{\hbar^2}{2m}\nabla^2\psi+V\psi+g\left(|\psi|^2-n_0\right)\psi
-\frac{mv_\infty^2}{2}\,\psi.
\label{eq:tdgp}
\end{equation}

The shedding simulation in the inset of Fig.~1 in the main manuscript starts from uniform flow in a domain of
$450\,\xi\times375\,\xi$, with the obstacle $125\,\xi$ from the inflow. Its inflow and lateral layers relax towards
the incoming wave, while a radiative condition is imposed on the outflow boundary. Vortex positions are determined by tracking cells with a phase winding $\pm2\pi$, which determines the circulation sign.

The obstacle radius is defined by equating the number of particles depleted by the Gaussian potential in the fluid at rest to the number that would be excluded by an equivalent hard disk, i.e. by solving $\pi a^2n_0=\int(n_0-|\psi_0|^2)\,d^2x$, where $\psi_0$ solves
Eq.~(1) with $v_\infty=0$.
This approach gives $a=4.29\,\xi$, corresponding to a density level of $0.6\,n_0$. Flowing
states with $v_\infty/c\le0.063$ give the same value within $0.02\,\xi$.
\section{Sensitivity to the domain size}
\label{sec:domain}

We compare stationary solutions in the standard domain, $L_1=384\,\xi$, with solutions in the larger domain,
$L_2=512\,\xi$, keeping the obstacle, grid spacing, and boundary conditions fixed.

At $v_\infty/c=0.0606$, the lowest velocity in
Fig.~2 of the manuscript, the F\"oppl pair is farthest from the obstacle, with $r_0\approx24\,\xi$. Enlarging the
domain moves the upper vortex by $0.04\,\xi$, corresponding to a relative $r_0$ error of $0.15\%$. For
$v_\infty/c\ge0.48$, the displacement is below $0.002\,\xi$. The corresponding changes in the critical velocities are
$v_{c1}(L_2)-v_{c1}(L_1)=-4\times10^{-5}\,c$ and
$v_{c2}(L_2)-v_{c2}(L_1)=-6\times10^{-5}\,c$.

Table~\ref{tab:domain} gives the change in the positive real eigenvalue
$\sigma$ on the F\"oppl branch for some representative velocities $v_\infty >v_{c1}$.
After the fold, on the two-pair branch the antisymmetric eigenvalue changes by $+0.03\%$. These small changes are consistent
with the localization of the unstable mode away from the boundaries.

\begin{table}[htb]
\caption{Domain-size dependence of the antisymmetric growth rate on the F\"oppl branch. The relative error is computed as
$[\sigma(L_2)-\sigma(L_1)]/\sigma(L_1)$.}
\label{tab:domain}
\begin{ruledtabular}
\begin{tabular}{cc}
$v_\infty/c$ & Relative change in $\sigma$ \\
\hline
0.479 & $+0.01\%$ \\
0.504 & $+0.01\%$ \\
0.508 & $-0.11\%$ \\
\end{tabular}
\end{ruledtabular}
\end{table}

\section{F\"oppl's point-vortex theory}
\label{sec:pv}

F\"oppl introduced a two-dimensional model for an ideal fluid, i.e. an inviscid, incompressible fluid of uniform density, flowing past a solid obstacle in the presence of two discrete vortices as a schematic description of the wake produced by
a cylinder in a stream~\cite{Foppl13}. We recall here its derivation and the linear stability of its stationary
solutions.

\subsection*{Complex potential and method of images}

Consider an inviscid and incompressible flow around a cylinder of radius $a$ centered at the origin, with inlet velocity
$v_\infty$ along $x$, in the presence of two point vortices with circulations $\kappa_1=-\Gamma$ and
$\kappa_2=+\Gamma$, with $\Gamma>0$, at positions $z_1(t)$ and $z_2(t)$, where $z=x+iy$. We suppose that the vorticity vanishes
everywhere except at the two vortices, where it is concentrated,
$\omega=\sum_j\kappa_j\,\delta(\mathbf r-\mathbf r_j)$. By the theorems of Kelvin and Helmholtz, the circulations
$\kappa_j$ are then constant in time, and the vortices move with the fluid. Since the obstacle is
impermeable, the normal velocity vanishes on its surface, while the fluid slips freely along it, and the circulation
around it is zero. Outside the vortices the flow is irrotational and divergence-free, and is described by the complex
potential $w(z)=\Phi+i\Psi$, where $\Phi$ is the velocity potential and $\Psi$ the stream
function. The potential is obtained by superposing the potentials of the flows
generated separately by the stream and by each vortex, each of which follows from the method of images~\cite{MT68}, that we briefly recall in the following. On the circle $|z|=a$ one has $z\bar z=a^2$, that is,
$\bar z=a^2/z$. Let $f(z)$ be an analytic function. On the circle,
$\bar f(a^2/z)=\bar f(\bar z)=\overline{f(z)}$. The sum $f(z)+\bar f(a^2/z)$ therefore has zero imaginary part on the
obstacle surface, which is then a streamline. The function
$\bar f(a^2/z)$ moves each singularity $z_j$ of $f$ to its mirror image with respect to the circle $a^2/\bar z_j$.

Applying the method to the potential $v_\infty z$ of the unperturbed stream gives the potential of the free flow
in presence of the cylinder,
\begin{equation}
w_0=v_\infty\left(z+\frac{a^2}{z}\right).
\label{eq:w0}
\end{equation}
Applying it to the potential $(\kappa_j/2\pi i)\ln(z-z_j)$ of a point vortex at $z_j$ gives instead the potential of
a vortex of circulation $\kappa_j$ at $z_j$ in presence of the cylinder,
\begin{equation}
w_j=\frac{\kappa_j}{2\pi i}\ln\frac{z-z_j}{a^2/z-\bar z_j}.
\label{eq:wj}
\end{equation}
Since $a^2/z-\bar z_j=-(\bar z_j/z)(z-a^2/\bar z_j)$, Eq.~\eqref{eq:wj} contains, besides the vortex itself, an image
vortex of circulation $-\kappa_j$ at $a^2/\bar z_j$, inside the cylinder, and one of circulation $+\kappa_j$ at its
center. Adding Eq.~\eqref{eq:w0} and Eq.~\eqref{eq:wj} for each of the two vortices, the central images cancel because
$\kappa_1+\kappa_2=0$, and, up to an additive constant, the total potential is~\cite{Foppl13,MT68,VMS11}
\begin{equation}
w(z)=v_\infty\left(z+\frac{a^2}{z}\right)
+\frac{i\Gamma}{2\pi}\ln\frac{(z-z_1)(z-a^2/\bar z_2)}{(z-a^2/\bar z_1)(z-z_2)}.
\label{eq:pvpotential}
\end{equation}
For $z_2=\bar z_1$,
Eq.~\eqref{eq:pvpotential} reduces to Eq.~(2) of the main text. The complex velocity $u-iv$ is the
derivative of the potential,
\begin{equation}
u-iv=\frac{dw}{dz}=v_\infty\left(1-\frac{a^2}{z^2}\right)
+\frac{i\Gamma}{2\pi}\left(\frac{1}{z-z_1}-\frac{1}{z-a^2/\bar z_1}+\frac{1}{z-a^2/\bar z_2}-\frac{1}{z-z_2}\right).
\label{eq:pvvelocity}
\end{equation}

\subsection*{Stationary solutions behind the cylinder}

The dynamics of the problem follows by imposing that each vortex moves with
the velocity induced at its position by the rest of the flow. When Eq.~\eqref{eq:pvvelocity} is evaluated at $z_1$, the term $1/(z-z_1)$ must be omitted.
 This is a typical approach for point vortices, as they produce a velocity field that is singular at their own positions,
 while the velocity induced by the image vortices must be retained. Denoting by $W_j$ the
regular part of Eq.~\eqref{eq:pvvelocity} at $z_j$,
\begin{equation}
\begin{aligned}
W_1&=v_\infty\left(1-\frac{a^2}{z_1^2}\right)
-\frac{i\Gamma}{2\pi}\left(\frac{1}{z_1-z_2}+\frac{1}{z_1-a^2/\bar z_1}-\frac{1}{z_1-a^2/\bar z_2}\right),\\
W_2&=v_\infty\left(1-\frac{a^2}{z_2^2}\right)
+\frac{i\Gamma}{2\pi}\left(\frac{1}{z_2-z_1}-\frac{1}{z_2-a^2/\bar z_1}+\frac{1}{z_2-a^2/\bar z_2}\right),
\end{aligned}
\label{eq:Wj}
\end{equation}
the equations of motion of the vortices read
\begin{equation}
\dot z_1=\overline{W}_1,\qquad \dot z_2=\overline{ W}_2.
\label{eq:pvode}
\end{equation}

We will now concentrate on solutions with $z_2=\bar z_1$.
The velocity of the upper vortex thus reads
\begin{equation}
W_1=u_1-iv_1=v_\infty\left(1-\frac{a^2}{z_1^2}\right)
-\frac{i\Gamma}{2\pi}\left[\frac{1}{z_1-\bar z_1}
+\frac{1}{z_1-a^2/\bar z_1}-\frac{1}{z_1-a^2/z_1}\right].
\label{eq:vel}
\end{equation}
Define $z_1=x_0+iy_0$, $y_0>0$, $r_0=|z_1|$ and $D=|z_1^2-a^2|^2=(r_0^2-a^2)^2+4a^2y_0^2$. The transverse velocity is
\begin{equation}
v_1=\frac{2a^2x_0y_0\left[\Gamma y_0r_0^4-\pi v_\infty(r_0^2-a^2)D\right]}
{\pi r_0^4(r_0^2-a^2)D}.
\label{eq:v1}
\end{equation}
For a pair behind the cylinder, $x_0>0$, the condition $v_1=0$ gives
\begin{equation}
\Gamma y_0r_0^4=\pi v_\infty(r_0^2-a^2)D.
\label{eq:transverse}
\end{equation}
Eliminating $\Gamma$ from the streamwise condition $u_1=0$ results in
\begin{equation}
(r_0^2-a^2)^2-4y_0^2r_0^2
=(r_0^2-a^2-2y_0r_0)(r_0^2-a^2+2y_0r_0)=0.
\label{eq:curve}
\end{equation}
Since $r_0>a$ and $y_0>0$, the second factor is positive, and the first must vanish. On this curve
$D=4y_0^2(r_0^2+a^2)$, and Eq.~\eqref{eq:transverse} links the circulation and the position of the vortices. The solution $z_1$ of $W_1=0$ thus depends on the vortex strength $\Gamma$, which has been left undetermined, and as
$\Gamma$ varies it describes a curve in the $(x,y)$ plane. In polar coordinates, $z_1=r_0e^{i\vartheta_0}$, the
equilibrium is given by
\begin{equation}
2\sin\vartheta_0=1-\frac{a^2}{r_0^2},\qquad
\frac{\Gamma}{v_\infty a}=2\pi\left(R-\frac1R\right)\left(1-\frac1{R^4}\right),\qquad R=\frac{r_0}{a}>1,
\label{eq:fopplsm}
\end{equation}
which is equivalent to Eq.~(3) of the main text.
 The first relation defines F\"oppl's curve. It starts at the rear stagnation
point, $r_0=a$ and $\vartheta_0=0$, and approaches $\vartheta_0=30^\circ$ far from the cylinder. The equations also
admit the reflected branch $x_0<0$, with the pair upstream. The right-hand side of the second relation increases
monotonically from zero to infinity, so each value of $\Gamma/v_\infty a$ selects a unique equilibrium on the
downstream branch, with $x_0=r_0\cos\vartheta_0$ and $y_0=r_0\sin\vartheta_0$.

In an inviscid analysis such as the present one, the circulation $\Gamma$ is a conserved parameter set by the initial condition, since the vorticity introduced at the initial time remains unchanged throughout the subsequent evolution. This is not what happens in a
viscous fluid, where a well-defined wake is established after a sufficiently long time, independently of the
initial conditions.
In a quantum fluid, the circulation is not a free parameter. A singly quantized vortex has
$\Gamma=2\pi\hbar/m$, and Eq.~\eqref{eq:fopplsm} becomes
\begin{equation}
y_0\left(1-\frac{a^4}{r_0^4}\right)=\frac{\hbar}{2mv_\infty},
\label{eq:quantizedsm}
\end{equation}
so that the flow velocity alone fixes the position of the pair on F\"oppl's curve.

\subsection*{Stationary solutions beside the cylinder}

For $x_0=0$, write $z_1=iy_0$ with $y_0>a$. The transverse velocity vanishes identically, and the remaining
condition $u_1=0$ yields
\begin{equation}
\Gamma=4\pi v_\infty\,
\frac{(y_0^2+a^2)^2(y_0^2-a^2)}{y_0(y_0^4+4a^2y_0^2-a^4)}.
\label{eq:side}
\end{equation}
With quantized circulation,
the pair approaches the cylinder as the inlet velocity increases, starting from
$y_0\simeq\hbar/(2mv_\infty)$ at low velocity.
While discarded by F\"oppl as experimentally unobserved in the classical
case, this family of solutions quantitatively describes the GP side-pair branch, as shown in
Fig.~2(a) in the main manuscript.

\subsection*{Linear stability}

To study the local stability of the solution behind the cylinder, we give a small perturbation to the state of the
system, represented by the two vortex positions, and linearize Eqs.~\eqref{eq:pvode} with respect to it. Setting
$z_1=z_\mathrm{eq}+\delta z_1$ and $z_2=\bar z_\mathrm{eq}+\delta z_2$, the linearized equations are
\begin{subequations}
\label{eq:lin}
\begin{align}
\frac{d\delta z_1}{dt}&=\overline{\left[\frac{\partial W_1}{\partial z_1}\delta z_1
+\frac{\partial W_1}{\partial\bar z_1}\delta\bar z_1
+\frac{\partial W_1}{\partial z_2}\delta z_2
+\frac{\partial W_1}{\partial\bar z_2}\delta\bar z_2\right]},\label{eq:lina}\\
\frac{d\delta z_2}{dt}&=\overline{\left[\frac{\partial W_2}{\partial z_1}\delta z_1
+\frac{\partial W_2}{\partial\bar z_1}\delta\bar z_1
+\frac{\partial W_2}{\partial z_2}\delta z_2
+\frac{\partial W_2}{\partial\bar z_2}\delta\bar z_2\right]},\label{eq:linb}
\end{align}
\end{subequations}
with the derivatives evaluated at the equilibrium.
While the equilibrium position of $z_2$ is the conjugate of that of
$z_1$, the perturbations $\delta z_1$ and $\delta z_2$ are arbitrary and in general are not the conjugate of each
other. As a consequence of $z_2=\bar z_1$, however,
\begin{equation}
\frac{\partial W_2}{\partial z_2}=\overline{\frac{\partial W_1}{\partial z_1}},\qquad
\frac{\partial W_2}{\partial\bar z_2}=\overline{\frac{\partial W_1}{\partial\bar z_1}},\qquad
\frac{\partial W_2}{\partial z_1}=\overline{\frac{\partial W_1}{\partial z_2}},\qquad
\frac{\partial W_2}{\partial\bar z_1}=\overline{\frac{\partial W_1}{\partial\bar z_2}}.
\label{eq:symderiv}
\end{equation}
Taking the sum and the difference of Eq.~\eqref{eq:lina} and of the conjugate of Eq.~\eqref{eq:linb} then gives two
decoupled equations for $s_\pm=\delta z_1\pm\delta\bar z_2$,
\begin{equation}
\frac{ds_\pm}{dt}=\bar A_\pm\,\bar s_\pm+\bar B_\pm\,s_\pm,\qquad
A_\pm=\frac{\partial W_1}{\partial z_1}\pm\frac{\partial W_1}{\partial\bar z_2},\qquad
B_\pm=\frac{\partial W_1}{\partial\bar z_1}\pm\frac{\partial W_1}{\partial z_2}.
\label{eq:decoupled}
\end{equation}
Equations~\eqref{eq:decoupled} govern separately the evolution of symmetric perturbations, for which $s_-=0$, and
antisymmetric perturbations, for which $s_+=0$.
Symmetric perturbations change the downstream position and the separation of the pair, while
antisymmetric perturbations tilt the pair and shift it across the centerline.
Evaluating the derivatives of Eq.~\eqref{eq:Wj} at the equilibrium gives
\begin{align}
A_\pm&=\frac{2v_\infty a^2}{z_1^3}
+\frac{i\Gamma}{2\pi}\left[-\frac{1}{4y_0^2}
+\frac{\bar z_1^2}{(r_0^2-a^2)^2}
-\frac{z_1^2\pm a^2}{(z_1^2-a^2)^2}\right],\nonumber\\
B_\pm&=\frac{i\Gamma}{2\pi}\left[\frac{a^2}{(r_0^2-a^2)^2}\pm\frac{1}{4y_0^2}\right].
\label{eq:AB}
\end{align}
Each of Eqs.~\eqref{eq:decoupled} represents a system of two real equations. Writing $\delta z_1=\delta x+i\delta y$, $A_\pm=\alpha_\pm+i\beta_\pm$
and $B_\pm=i\gamma_\pm$, it reads
\begin{equation}
\frac{d}{dt}\begin{pmatrix}\delta x\\\delta y\end{pmatrix}
=\mathsf M_\pm\begin{pmatrix}\delta x\\\delta y\end{pmatrix},\qquad
\mathsf M_\pm=\begin{pmatrix}
\alpha_\pm&\gamma_\pm-\beta_\pm\\
-\gamma_\pm-\beta_\pm&-\alpha_\pm
\end{pmatrix}.
\label{eq:pvmatrix}
\end{equation}
The ansatz
\begin{equation}
\begin{pmatrix}\delta x\\\delta y\end{pmatrix}=\mathbf q\,e^{\lambda t},
\label{eq:ansatz}
\end{equation}
with $\mathbf q$ a constant vector, leads to the eigenvalue problem
$\mathsf M_\pm\mathbf q=\lambda\mathbf q$, that has solutions
\begin{equation}
\lambda_\pm^2=-\det\mathsf M_\pm=|A_\pm|^2-\gamma_\pm^2.
\label{eq:lambda}
\end{equation}
On F\"oppl's curve, the calculation gives~\cite{VMS11}
\begin{align}
-\lambda_+^2=\omega_\mathrm{PV}^2
&=\frac{v_\infty^2}{a^2}\frac{3R^6+5R^4+13R^2-5}{R^{10}},\nonumber\\
\lambda_-^2=\sigma_\mathrm{PV}^2
&=\frac{v_\infty^2}{a^2}\frac{3R^6+3R^4-3R^2+1}{R^{10}},\qquad R=\frac{r_0}{a}>1.
\label{eq:pvrates}
\end{align}
Both right-hand sides are positive. The eigenvalues of the symmetric equation are therefore imaginary,
$\lambda_+ = \pm i\omega_\mathrm{PV}$, while those of the antisymmetric equation are real, $\lambda_-=\pm\sigma_\mathrm{PV}$. F\"oppl's
solution, for any $\Gamma$, is thus neutrally stable with respect to symmetric perturbations, which oscillate about
the equilibrium without decaying, and unstable with respect to antisymmetric perturbations, with one growing and one
decaying direction. The full equilibrium is a saddle-center, and no value of $\Gamma$ therefore makes F\"oppl's solution stable.
The eigenvector associated with $\sigma_\mathrm{PV}$ is computed as $\mathsf M_-\mathbf q=\sigma_\mathrm{PV}\mathbf q$.

\section{Bogoliubov modes}
\label{sec:bdg}

Writing a small perturbation of the stationary state as
$\psi=\psi_0+u\,e^{\lambda t}+\bar v\,e^{\bar\lambda t}$ gives
\begin{equation}
i\hbar\lambda\begin{pmatrix}u\\v\end{pmatrix}
=\begin{pmatrix}\mathcal L&g\psi_0^2\\-g\bar\psi_0^2&-\bar{\mathcal L}\end{pmatrix}
\begin{pmatrix}u\\v\end{pmatrix},\qquad
\mathcal L=-\frac{\hbar^2}{2m}\nabla^2+V+g(2|\psi_0|^2-n_0)-\frac{mv_\infty^2}{2}.
\label{eq:bdg}
\end{equation}
Here $u$ and $v$ are perturbations of small amplitude and $\operatorname{Re}\lambda>0$ corresponds to unstable modes.
The eigenproblem uses the same spatial discretization
as the stationary calculation and homogeneous boundary conditions for the perturbations~\cite{GVLC26b}.
At every velocity considered in the manuscript, we find a real
pair $\pm\sigma$ (Table~\ref{tab:sigma}) associated to a mode that is antisymmetric under $y\to -y$, and a symmetric conjugate pair $\lambda_s$, $\overline{\lambda}_s$ (Table~\ref{tab:sym}). The directions of the GP and point-vortex
core displacements are in good agreement, as shown in Fig.~3(c) in the main manuscript for some reference velocities, while Table~\ref{tab:sigma} lists the antisymmetric GP growth rate and the point-vortex prediction for $a=4.29\,\xi$ and
$\Gamma=2\pi\hbar/m$. The error can be reduced using an
effective radius and circulation determined from the measured equilibrium position $(x_0,y_0)$:
\begin{equation}
a_\mathrm{eff}=\sqrt{r_0(r_0-2y_0)},\qquad
\Gamma_\mathrm{eff}=4\pi v_\infty y_0\left(1-\frac{a_\mathrm{eff}^4}{r_0^4}\right).
\label{eq:effective}
\end{equation}
The resulting rate $\sigma_\mathrm{PV,eff}$ is then evaluated from point-vortex theory with the effective parameters.

\begin{table}[htb]
\caption{Antisymmetric growth rate on the F\"oppl branch ($L=384\,\xi$). $r_0/a$ is the point-vortex position for
$a=4.29\,\xi$ and $\Gamma=2\pi\hbar/m$.}
\label{tab:sigma}
\begin{ruledtabular}
\begin{tabular}{ccccccc}
$v_\infty/c$ & $r_0/a$ (PV) & $\sigma$ $[10^{-2}\,c/\xi]$ & $\sigma/\sigma_\mathrm{PV}$ & $a_\mathrm{eff}/\xi$ &
$\Gamma_\mathrm{eff}m/2\pi\hbar$ & $\sigma/\sigma_\mathrm{PV,eff}$\\
\hline
0.100 & 3.59 & 0.314 & 0.968 & 4.267 & 1.005 & 0.980 \\
0.150 & 2.63 & 0.907 & 0.970 & 4.283 & 1.007 & 0.981 \\
0.201 & 2.18 & 1.77 & 0.959 & 4.340 & 1.020 & 0.980 \\
0.249 & 1.94 & 2.74 & 0.939 & 4.385 & 1.036 & 0.973 \\
0.298 & 1.79 & 3.81 & 0.915 & 4.413 & 1.055 & 0.963 \\
0.354 & 1.67 & 5.08 & 0.888 & 4.429 & 1.077 & 0.948 \\
0.398 & 1.60 & 6.10 & 0.867 & 4.434 & 1.094 & 0.934 \\
0.452 & 1.54 & 7.33 & 0.842 & 4.437 & 1.113 & 0.913 \\
0.479 & 1.51 & 7.92 & 0.829 & 4.431 & 1.124 & 0.903 \\
0.499 & 1.49 & 8.35 & 0.820 & 4.418 & 1.138 & 0.897 \\
0.504 & 1.49 & 8.47 & 0.818 & 4.409 & 1.145 & 0.897 \\
0.509 & 1.49 & 8.61 & 0.819 & 4.386 & 1.162 & 0.904 \\
\end{tabular}
\end{ruledtabular}
\end{table}

\begin{table}[htb]
\caption{Symmetric oscillatory eigenvalue $\lambda_s$ on the F\"oppl branch ($L=384\,\xi$). $\omega_\mathrm{PV}$ is the
symmetric frequency of F\"oppl's model at the same velocity ($a=4.29\,\xi$, $\Gamma=2\pi\hbar/m$).}
\label{tab:sym}
\begin{ruledtabular}
\begin{tabular}{cccc}
$v_\infty/c$ & $\mathrm{Re}\,\lambda_s$ $[10^{-3}\,c/\xi]$ & $\mathrm{Im}\,\lambda_s$ $[c/\xi]$ & $\mathrm{Im}\,\lambda_s/\omega_\mathrm{PV}-1$ $[\%]$\\
\hline
0.5066 & 6.20 & 0.1273 & $-12.5$ \\
0.5040 & 6.23 & 0.1265 & $-12.2$ \\
0.4988 & 6.19 & 0.1248 & $-11.7$ \\
0.4961 & 6.14 & 0.1240 & $-11.5$ \\
0.4906 & 6.01 & 0.1222 & $-11.0$ \\
0.4878 & 5.93 & 0.1212 & $-10.8$ \\
0.4792 & 5.67 & 0.1183 & $-10.1$ \\
0.4745 & 5.50 & 0.1167 & $-9.8$ \\
0.4697 & 5.37 & 0.1150 & $-9.4$ \\
0.4648 & 5.22 & 0.1132 & $-9.1$ \\
0.4599 & 5.05 & 0.1115 & $-8.7$ \\
0.4549 & 4.91 & 0.1096 & $-8.4$ \\
0.4524 & 4.84 & 0.1087 & $-8.3$ \\
\end{tabular}
\end{ruledtabular}
\end{table}

\section{Ghost vortices inside the obstacle}
\label{sec:ghost}

\begin{figure}[t]
\centering
\includegraphics[width=\columnwidth]{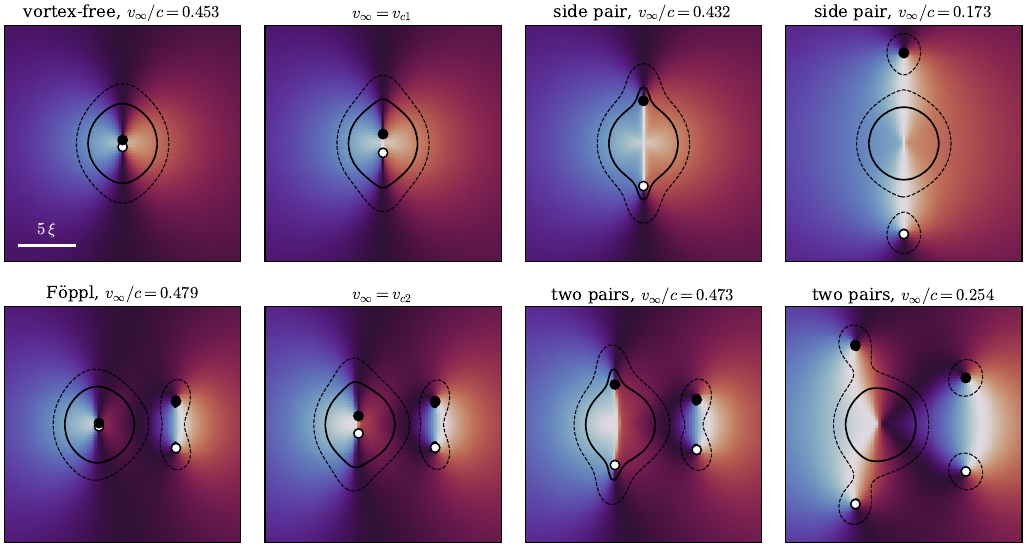}
\caption{Phase of $\psi e^{-imxv_\infty/\hbar}$ revealing ghost vortices at the two folds. Solid and dashed lines: density isocontours $|\psi|^2=0.05\,n_0$ and $0.5\,n_0$. Black and
white circles: zeros of negative and positive circulation. Top row: from the vortex-free to the side-pair branch
through $v_{c1}$. Bottom row: from the F\"oppl to the two-pair branch through $v_{c2}$.}
\label{fig:ghost}
\end{figure}

Inside the obstacle the density is exponentially small. Nevertheless, the phase of $\psi$ can still be defined and its singularities can be
located by phase-winding. Near the folds, $\psi$ has a pair of zeros of opposite circulation inside the
obstacle, at $(x_g,\pm y_g)$, with the negative-circulation zero in the upper half-plane. On the vortex-free branch, this happens for $v_\infty/c\gtrsim0.39$, with $x_g=0$, and on the F\"oppl branch for
$v_\infty/c\gtrsim0.46$, with $|x_g|<0.13\,\xi$. At lower velocities the two zeros merge on the axis and disappear. As
the velocity increases towards the fold, the zeros move apart and they continue to move outwards as the
velocity decreases after the critical value, crossing $r=a$ at $v_\infty/c\approx0.41$ on the side-pair branch and $0.45$ on the two-pair branch.
After the emission, they become the side vortices of the side-pair branch and the second pair of the two-pair branch, respectively
(Fig.~\ref{fig:ghost}).
The two folds also share their spectral structure, as at both critical velocities a real symmetric eigenvalue $\sigma_s$ vanishes. 

\section{Supplemental Movie 1: controlled wake and onset of shedding}
\label{sec:movie}

Supplemental Movie~1 shows the density $n/n_0$ at $v_\infty=0.479\,c$ in the
$40\,\xi\times24\,\xi$ near-wake window of Fig.~4 of the main manuscript. The flow is from left to right, and crosses mark the stationary core positions.
The accompanying curve shows the largest distance $\delta$ of the two vortex
cores from their stationary positions relative to the obstacle. The first part compares the uncontrolled and
controlled flows after the same initial obstacle displacement, up to $\sigma t=25$. Without feedback, $\delta$ reaches
$\xi$ at $\sigma t\simeq9$. The second part shows the controlled state up to the moment when feedback is switched off at $\sigma t=105$. The vortex pair subsequently escapes and vortex shedding resumes. The corresponding curves show the vortex displacement $\delta$ and the controlled velocity magnitude $|\dot{\mathbf{X}}|/c$. After the release, the pair initially tilts in the direction of the antisymmetric mode shown in
Fig.~3(c) of the main manuscript and escapes along an inclined trajectory, suggesting a connection with the symmetry
breaking in V-shaped shedding observed in the uncontrolled dynamics.

\FloatBarrier
\bibliography{bibliomain}